\documentclass{elsart}
\usepackage{latexsym}

\newtheorem{theorem}{Theorem}

\newtheorem{condition}[theorem]{Condition}

\newtheorem{corollary}[theorem]{Corollary}

\newtheorem{definition}[theorem]{Definition}

\newtheorem{lemma}[theorem]{Lemma}

\newtheorem{proposition}[theorem]{Proposition}

\newcommand{\be}{\begin{equation}}
\newcommand{\ee}{\end{equation}}
\newcommand{\bea}{\begin{eqnarray}}
\newcommand{\eea}{\end{eqnarray}}

\begin{document}
\begin{frontmatter}
\title{New properties of Cauchy and event horizons}
\author{Robert J. Budzy\'nski}
\address{Department of Physics,
Warsaw University,
Ho\.za 69,
00-681 Warsaw, Poland} 
\author{Witold Kondracki}
\author{Andrzej Kr\'olak}
\address{Institute of Mathematics,
Polish Academy of Sciences,
\'Sniadeckich 8,
00-950 Warsaw, Poland}

\begin{abstract}
We present several recent results concerning Cauchy and event horizons.
In the first part of the paper we review the differentiablity properties
of the Cauchy and the event horizons. In the second part 
we discuss compact Cauchy horizons and summarize their main properties.
\end{abstract}
%\PACS 53C50 \sep 53C80 \sep 83C75
\end{frontmatter}
\section{Introduction}

Cauchy horizons and event horizons play an important role in 
relativity. The consequence of causality which is one of the  basic postulates 
of relativity is that data on an intial surface $S$ determine the evolution of
the relativistic equations in the domain of dependence which is the set of points
$p$ such that all past-directed causal curves from $p$ intersect $S$.
The future boundary of domain of dependence, if non-empty, is called the Cauchy
horizon and it is a null surface. To the future of a Cauchy horizon space-time
cannot be predicted from the initial surface $S$. A fundamental unresolved
problem in classical relativity posed by Roger Penrose is whether there
is a "cosmic censor" that ensures existence of an initial surface from which 
the whole of space-time is predictable and a Cauchy horizon does not occur.
Hence it is of interest to study the properties of Cauchy horizons and conditions
on space-times under which they can or cannot arise.
The event horizon is the boundary of the region of space-time to the past of the
boundary at infinity. The event horizon like a Cauchy horizon 
is also a null surface. The region of space-time to the future of the event horizon
is the black hole region from which observers cannot escape and which may contain a
space-time singularity at which curvature grows unboundedly. This indicates
breakdown of classical general relativity theory. No satisfactory theory exists
till now to describe these singular regions. Black hole regions should arise
as a result of the collapse of stars and galaxies and the astronomical observations
show that almost certainly this is the case. Thus study of the properties
of event horizon has attracted great interest.

\section{Preliminaries}

In this Section we shall recall several basic definitions
and properties, all of which can be found, e.g., in \cite{HE73,BEE96}.

\begin{definition}
A space-time $(M,g)$ is a separable smooth $n-$dimensional, Hausdorff manifold $M$
with a pseudo-Riemannian metric $g$ of signature $(-,+,...,+)$ and a time orientation.
\end{definition}

Let $I^+(S,U), I^-(S,U)$ be respectively chronological future and chronological past
of set $S$ relative to set $U$ and let $J^+(S,U), J^-(S,U)$ be the relative
causal future and past. When the set $U$ is omitted the chronological and causal
pasts and futures are relative to space-time manifold.

A set $\mathcal{S}$ is said to be {\em achronal} respectively {\em acausal} 
if there are no two points of $\mathcal{S}$ with timelike respectively causal separation.
The {\em edge}, edge($A$), of the achronal set $A$ consists of all points $p$
in $\overline{A}$ such that every neighborhood $U$ of $p$ contains a timelike curve 
from $I^-(p,U)$ to $I^+(p,U)$
which does not meet $A$. The set $A$ is said to be edgeless if $edge(A) = \emptyset$. 

\begin{definition}
A {\em partial Cauchy surface} $\mathcal{S}$ is a connected, acausal, edgeless $n-1$
dimensional submanifold of $(M,g)$.
\end{definition}

We give definitions and state our results in terms of the future horizon 
$H^+(\mathcal{S})$, but similar results hold for any past Cauchy horizon $H^-(\mathcal{S})$.

\begin{definition}
The {\em future Cauchy development} $D^{+}(\mathcal{S})$ consists of all points $p\in M$ 
such that each past endless and past directed causal curve from $p$
intersects the set $\mathcal{S}$. The {\em future Cauchy horizon} is 
$H^{+}(\mathcal{S})=\overline{(D^{+}(\mathcal{S}))}-I^{-}(D^{+}(\mathcal{S}))$.
\end{definition}

The future Cauchy horizon is generated by null geodesic segments.
Let $p$ be a point of the Cauchy horizon; then there is at least one null
generator of $H^{+}(\mathcal{S})$ containing $p$. When a null generator of $H^{+}(\mathcal{S})$ 
is extended into the past it either has no past endpoint 
or has a past endpoint on $edge(\mathcal{S})$ [see \cite{HE73}, p. 203]. 
However, if a null generator is extended
into the future it may have a last point on the horizon which then said to
be an {\em endpoint} of the horizon. We define the {\em multiplicity} [see 
\cite{BK98}] of a point $p$ in $H^{+}(\mathcal{S})$ to be the number of null
generators containing $p$. Points of the horizon which are not endpoints
must have multiplicity one. The multiplicity of an endpoint may be any
positive integer or infinite. We call the set of endpoints of multiplicity
two or higher the {\em crease set}, compare \cite{CG98}. 

Let the space­time $(M,g)$ be strongly future asymptotically predictable 
and let ${\mathcal J}^{+}$ be the future null infinity (see \cite{HE73} Chapter 9).
The set $B = M - J^-({\mathcal J}^{+})$ is called the black hole region and  
E = $\dot{J}^{-}({\mathcal J}^{+})$ is the {\em event horizon}. The event horizon like a
Cauchy horizon is generated by null geodesic segments however the generators of
the event horizon have no future endpoints but may have past endpoints.

\section{Non-differentiable Cauchy and event horizons}

By a basic Proposition due to Penrose [\cite{HE73}, Prop. 6.3.1] $H^{+}(\mathcal{S})$ 
is an $n-1$ dimensional Lipschitz topological submanifold of $M$ and is achronal. 
Since a Cauchy horizon is Lipschitz it follows from a theorem of Rademacher that
it is differentiable almost everywhere (i.e. differentiable except for a set
of $n-1$ dimensional measure zero). This does not exclude the possibility
that the set of non-differentiable points is a dense subset of the horizon.
In this section we shall give examples of such a behaviour.

$H^{+}(\mathcal{S})$ is differentiable if it is a differentiable submanifold of $M$.
Thus if $H^{+}(\mathcal{S})$ is differentiable at the point $p$, then there is a well
defined 3-dimensional linear subspace $N_0$ in the tangent space $T_{p}(M)$
such that $N_{0}$ is tangent to the 3-dimensional surface $H^{+}(\mathcal{S})$ at $p$.

\begin{theorem}
(Chru\'sciel and Galloway \cite{CG98})

There exists a connected set $\mathcal{U} \subset R^2 = \{t = 0\} \subset R^{2,1}$,
where $R^{2,1}$ is a $2 + 1$ dimensional Minkowski space-time, with the
following properties:

\begin{enumerate}
\item  The boundary $\partial \mathcal{U} = \bar{\mathcal{U}} - {\rm int}\, 
\mathcal{U}$ of $\mathcal{U}$ is a
connected, compact, Lipschitz topological submanifold of $R^2$. $\mathcal{U}$ is the
complement of a compact set in $R^2$.

\item  There exists no open set $\Omega \subset R^{2,1}$ such that $\Omega
\cap H^+(\mathcal{U})$ is a differentiable submanifold of $R^{2,1}$.
\end{enumerate}
\end{theorem}

A further study has shown that the densely nondifferentiable Cauchy horizons 
are quite common \cite{BKK99}. Let $R^{2,1}$ be the $3$
-di\-men\-sion\-al Minkowski space-time. Let $\Sigma$ be the
surface ${t = 0}$, and let $K$ be a compact, convex subset of $\Sigma$. 
Let $\partial K$ denote the boundary of $K$. 

Let ${\mathcal H}$ be the set of Cauchy horizons arising from compact convex
sets $K \subset \Sigma$. The topology on ${\mathcal H}$ is induced by the
Hausdorff distance on the set of compact and convex regions K. 

\begin{theorem}
\label{Th:BKKnd}
Let ${\mathcal H}$ be the set of future Cauchy horizons $H^+(K)$ where $K$ are
compact and convex regions of $\Sigma$. The subset of densely
nondifferentiable horizons is dense in ${\mathcal H}$.
\end{theorem}

The above theorem generalizes to the 3-dimensional case. 

It also possible to construct examples of densely nondifferentiable Cauchy 
horizons of partial Cauchy surfaces and also the existence of densely 
nondifferentiable black hole event horizons.
\bigskip

\noindent {\bf Example 1:} {\em A rough wormhole.} \medskip

Let $R^{3,1}$ be the 4-dimensional Minkowski space-time and let $K$ be a
compact subset of the surface $\{t = 0\}$ such that its Cauchy horizon is
nowhere differentiable in the sense of Theorem \ref{Th:BKKnd}.
We consider a space-time obtained by removing the complement of the interior
of the set $K$ in the surface ${t = 0}$ from the Minkowski space-time. Let
us consider the partial Cauchy surface $\mathcal{S} = \{t = -1\}$. The future Cauchy
horizon of $\mathcal{S}$ is the future Cauchy horizon of set $K - {\rm edge}(K)$,
since ${\rm edge}(K)$ has been removed from the space-time. Thus the future
Cauchy horizon is nowhere differentiable and it is generated by past-endless
null geodesics. The interior of the set $K$ can be thought of as a
``wormhole'' that separates two ``worlds'', one in the past of surface $\{t
= 0\}$ and one in its future. \bigskip

\noindent {\bf Example 2:} {\em A transient black hole.} \medskip

Let $R^{3,1}$ be the 4-dimensional Minkowski space-time and let $K$ be a
compact subset of the surface $\{t = 0\}$ such that its {\em past} Cauchy
horizon is nowhere differentiable in the sense of Theorem \ref{Th:BKKnd}. 
We consider a space-time obtained by removing from Minkowski
space-time the closure of the set $K$ in the surface ${t = 0}$. Let us
consider the event horizon E = $\dot{J}^{-}({\mathcal J}^{+})$. The event
horizon $E$ coincides with $H^-(K)-{\rm edge}(K)$ and thus it is not empty
and nowhere differentiable. The event horizon disappears in the future of
surface $\{t = 0\}$ and thus we can think of the black hole (i.e. the set $B
:= R^{3,1} - J^{-}({\mathcal J}^{+})$) in the space-time as ``transient''.
\bigskip

The following results summarize differentiability properties 
of Cauchy horizons.

\begin{theorem} (Beem and Kr\'olak \cite{BK98}, Chruœciel \cite{Ch1998})
\label{Th:BKdif}
A Cauchy horizon is differentiable at all points of multiplicity one.
In particular, a Cauchy horizon is differentiable at an endpoint where
only one null generator leaves the horizon.
\end{theorem}

\begin{proposition} (Beem and Kr\'olak \cite{BK98})
\label{P:BKdif}
Let $W$ be an open subset of the Cauchy horizon $H^+(\mathcal{S})$. Then the following
are equivalent:
\begin{enumerate}
\item  $H^+(\mathcal{S})$ is of class $C^r$ on $W$ for some $r \geq 1$.
\item  $H^+(\mathcal{S})$ has no endpoints on $W$.
\item  All points of $W$ have multiplicity one.
\end{enumerate}
\end{proposition}

Note that the three parts of Proposition \ref{P:BKdif} 
are logically equivalent for an {\em open} set $W$, 
but that, in general, they are not equivalent for sets which fail to be open. 
Using the equivalence of parts
(1) and (3) of Proposition \ref{P:BKdif}, it now follows that near 
each endpoint of multiplicity one there must be points where the horizon 
fails to be differentiable. Hence, each neighborhood of 
an endpoint of multiplicity one must contain endpoints of higher 
multiplicity. This yields the following corollary.

\begin{corollary} (Beem and Kr\'olak \cite{BK98})

If $p$ is an endpoint of multiplicity one on a Cauchy horizon $H^+(\mathcal{S})$, then
each neighborhood $W(p)$ of $p$ on $H^+(\mathcal{S})$ contains points where the
horizon fails to be differentiable. Hence, the set of endpoints of
multiplicity one is in the closure of the crease set.
\end{corollary}

However the following conjecture remains so far unresolved.

{\bf Conjecture.}  The set of all endpoints
of a Cauchy horizon must have measure zero.

\section{Compact and compactly generated Cauchy horizons}
\label{CH:com}

In this section we shall assume that space-time manifold is 4-dimensional.
We shall say that a Cauchy horizon $H$ is {\bf almost smooth} if
it {\em contains an open set G where it is $C^2$ and such that 
complement of $G$ in $H$ has measure zero}. Throughout the rest of 
this section we shall assume that all Cauchy horizons are almost smooth.
At the end of this section we shall mention recent results
by which it may be posible to relax the above differentiability assumption.

Let $t$ be an affine parameter on a null geodesic $\lambda$, $k^a$ be components
of the tangent vector to $\lambda$, $\theta$ be expansion,
$\sigma$ be shear and $R_{ab}$ be components of the Ricci tensor.
Then we have the following ordinary non-linear equation of Riccati type for
$\theta$ .
\be
\frac{d\theta}{dt} = -\frac{1}{2}\theta^2 -
2\sigma^2 - R_{ab}k^ak^b , \label{eq:RNP}
\ee
The quantity $\theta$ describes the expansion 
of congruences of null geodesics 
infinitesimally neighboring $\lambda$ and it is defined as  
$\theta = \frac{1}{A}\frac{dA}{dt}$ where $A$ is cross-section
area of the congruence.
The above equation is used extensively in geometrical techniques to study 
the large-scale structure of space-time developed by Geroch, Hawking, 
and Penrose and we shall call it Raychaudhuri-Newman-Penrose (RNP) equation.

We shall first introduce the following two conditions.

\begin{condition}[Null convergence condition]
We say that the null convergence condition holds if $R_{ab}k^ak^b \geq 0$ 
for all null vectors $k$.
\label{c:conv}
\end{condition}

Let $R_{abcd}$ be components of Riemann tensor.
We say that an endless null geodesic $\gamma$ is
{\bf generic} if for some point $p$
on $\gamma$, $k^ck^dk_{[a}R_{b]cd[e}k_{f]} \neq 0$ where $k$ is a vector
tangent to $\gamma$ at $p$.

\begin{condition}[Generic condition]
All endless null geodesics in space-time are generic.
\label{c:gen}
\end{condition}

By the Einstein equations the null convergence condition means that the local energy
density is non-negative and it is satisfied by all reasonable classical matter models.
The generic condition means that every null geodesic encounters 
some matter or radiation that is not pure radiation moving in the direction of
the geodesic tangent.
The above two purely geometrical conditions have a very clear physical interpretation
and they are reasonable to impose for any classical matter fields like gravitational
and electromagnetic fields.

\begin{definition}
Let $\mathcal{S}$ be a partial Cauchy surface.
A future Cauchy horizon $H^+(\mathcal{S})$ is compactly generated if
all its generators, when followed into their past, 
enter and remain in a compact subset $C$.
\label{d:cgen}
\end{definition}

The above class of Cauchy horizons has been introduced by Hawking 
\cite{H92} to describe a situation in which a Cauchy horizon
arises as a result of causality violation rather than singularities or
timelike boundary at infinity.

{\em Remark:} It is clear that every compact Cauchy horizon is compactly generated
where the set $C$ in the above definition is the horizon itself.

\begin{theorem}
\label{Th:Hcg}
If the null convergence condition holds then a compactly generated Cauchy horizon 
that is non-compact cannot arise.
\end{theorem}

The proof of Theorem \ref{Th:Hcg} will be outlined later in the paper.

\begin{corollary}[Hawking 1992\cite{H92}]
\label{Co:Hcg}
If null convergence condition holds then a compactly generated Cauchy horizon 
$H^+(\mathcal{S})$ cannot arise from a non-compact partial Cauchy surface $\mathcal{S}$.
\end{corollary}

\emph{Outline of the proof:}
The result follows from Theorem \ref{Th:Hcg} because if a partial Cauchy surface
$\mathcal{S}$ is noncompact then the future Cauchy horizon $H^+(\mathcal{S})$ 
cannot be compact. $\Box$

Thus under a very mild - from physical point of view - restriction 
on space-time a nontrivial class of Cauchy horizons is ruled out.

For the case of compact Cauchy horizons we have the following result.

\begin{theorem}
\label{Th:Bch}
If null convergence condition holds and at least one of the null geodesic 
generators of a Cauchy horizon $H$ is generic then $H$ cannot be compact.
\end{theorem}

The proof of the above results relies on several lemmas which
summarize basic properties of compact and compactly generated 
Cauchy horizons. 

\begin{lemma}[Hawking and Ellis 1973\cite{HE73}]
\label{l:hcom}
Let $H^{+}\left( \mathcal{S}\right) $ be a  compact future
Cauchy horizon for a partial Cauchy surface $\mathcal{S}$, then the null
geodesic generating segments of $H^{+}\left( \mathcal{S}\right) $ are
geodesically complete in the past direction.
\end{lemma}

\emph{Outline of the proof:}
One shows that when a null generator $\gamma $ of the future Cauchy horizon
is past incomplete, $\gamma $ can be varied into the past to give a
past-inextendible timelike curve imprisoned in a compact subset of 
the interior of the future domain of dependence $D^{+}\left( \mathcal{S}\right) $. 
This is a contradiction as all timelike past-intextendible curves in 
$intD^{+}\left( \mathcal{S}\right) $ must intersect $\mathcal{S} $. $\Box$

\begin{lemma}[Hawking and Ellis 1973\cite{HE73}]
Let the null convergence condition hold.
Then the expansion $\theta$ and the shear $\sigma$ of null geodesic
generators of a compact Cauchy horizon $H^+(\mathcal{S})$ are zero.
\label{l:hexp}
\end{lemma}

\emph{Outline of the proof:}
On space-time manifold $M$ one can introduce a Riemannian positive
definite metric $\hat{g}$.
Let $G$ be the set defined at the beginning of the section. Through every point
of $G$ there passes a unique generator of $H^+(\mathcal{S})$.
Following Hawking we introduce a map
$u_t: G \rightarrow G$, see [\cite{H92}, p. 606] which moves each point
of $G$ a proper distance $t$ in the metric $\hat{g}$ into the past 
along a generator of $H^+(\mathcal{S})$.
We have the equation
\be
d/dt \int_{ u_t(G)}  dA = 2 \int_{ u_t(G)} \theta dA
\label{e:Hmap}
\ee
Notice that since $H^{+}(\mathcal{S})$ is compact we find that $\int_{ u_t(G)}
dA$  is finite.  The derivative of $ \int_{ u_t(G)}  dA$
cannot be positive since the set $G$ is mapped into itself.  Thus, the
right hand side of Equation \ref{e:Hmap} is nonpositive.  On the other hand 
$\theta \geq 0$ since otherwise by null convergence condition and past-completeness
from the RNP equation there would be a past endpoint on a generator of 
$H^{+}(\mathcal{S})$. Thus the right hand side of Equation \ref{e:Hmap} is nonnegative.
Since by null convergence condition and RNP equation $\theta$ 
is a monotonic function the only possibility is that
$\theta = 0$. Consequently from RNP equation and the null convergence condition 
it follows that $\sigma = 0$ as well. $\Box$

Let $\lambda : I \rightarrow M$ be a continuous curve defined on an open interval, 
$I \in R^1$, which may be infinite or semi-infinite.  
We say that point $x \in M$ is a {\em terminal accumulation point}
of $\lambda$ if for every open neighborhood $O$ of $x$ 
and every $t_o \in I$ there exists $t \in I$ with $t > t_o$ such that 
$\lambda(t) \in O$.
When $\lambda$ is a causal curve, we call $x$ a {\em past terminal accumulation point}
if it is a terminal accumulation point when $\lambda$ 
is parametrized so as to make it past-directed.

\begin{definition}
Let $H^+(\mathcal{S})$ be a compactly generated future Cauchy horizon. The {\em base set} $B$
is defined by
$B$ = \{ $x \in H^+(\mathcal{S})$ : there exists a null generator, $\lambda$ of
       $H^+(\mathcal{S})$ such that $x$ is a past terminal accumulation point of 
       $\lambda$ \}
\end{definition}

We have the following proposition.

\begin{proposition}[Kay et al.\cite{KRW97}]
The base set $B$ of any compactly generated Cauchy horizon, $H^+(\mathcal{S})$, always is nonempty.
In addition, all the generators of $H^+(\mathcal{S})$ asymptotically approach $B$ 
in the
sense that for each past-directed generator, $\lambda$, of $H^+(\mathcal{S})$ 
and each open 
neighbourhood $O$ of $B$, there exists a $t_o \in I$ (where $I$ is the interval 
of definition of $\lambda$) such that $\lambda(t) \in O$ for all $t > t_o$. 
Finally, $B$ is comprised by null geodesic generators, $\gamma$,
of $H^+(\mathcal{S})$ which are contained entirely within $B$ and are both past and
future inextendible.
\label{l:hend}
\end{proposition}

\emph{Outline of the proof:}
The first two properties follow easily from the definitions. 
To prove the last property one chooses a point $x \in B$ and considers 
a past inextendible generator $\lambda$ such that $x$ is a past terminal 
accumulation point of $\lambda$.
One takes a sequence of points $p_i$ on $\lambda$ and a sequence 
of tangent vectors $k_i$ to $\lambda$ at each $p_i$. By compactness of 
the Cauchy horizon there exists a tangent vector $k$ at $x$ such that 
$\{(p_i,k_i)\}$ converges to $\{(p,k)\}$. This determines an inextendible null 
geodesic $\gamma$ through $x$. One then shows that $\gamma$ 
is contained in $B$ and that it is inextendible
using the fact that for an arbitrary point $y$ on $\gamma$ 
there is a sequence of points on $\lambda$ converging to $y$. $\Box$

{\em Remark}. The proof that a compact Cauchy horizon must contain an
endless null generator has independently been given by Borde \cite{B1984}
and Hawking \cite{H92}.

\begin{lemma}[Beem and Kr\'olak 1998\cite{BK98}]
Let the null convergence condition hold. Then
all null geodesic generators of a compact future Cauchy horizon 
$H^{+}(\mathcal{S}) $ are endless.
\label{l:haend}
\end{lemma}

\emph{Outline of the proof:}
Assume that $H^{+}(\mathcal{S})$ has an endpoint $p$ of a null generator $\gamma$.
Even if $\gamma$ does
not lie in $G$, the horizon will be differentiable on the part of
$\gamma$ in the past of $p$.  Choose some
$q$ on $\gamma$ in the past of $p$ and some $u$ on $\gamma$ in the future
of $p$.
Then for some small neighborhood $W(q)$ of $q$ on $H^{+}(\mathcal{S})$, 
all null generators of the horizon through
points $r$ of $W(q)$ will have directions close to the direction of $\gamma$ at
$q$. Recall that given a compact domain set in the $t-$axis, geodesics
with close initial conditions remain close on the compact domain set.
Thus, by choosing $W(q)$ sufficiently small all null
generators through points of $W(q)$ will come arbitrarily close to $u$.  Therefore
for sufficiently small $W(q)$ all null generators intersecting $W(q)$ must
leave $H^{+}(\mathcal{S})$ in the future.  Since $W(q)$ is open in 
$H^{+}(\mathcal{S})$ it must
have a nontrivial intersection with the set $G$.  Thus, $u_t(G)$ cannot be all of $G$
for some positive values of $t$ and this yields a negative for some values of
$t$ on the left hand side of Equation \ref{e:Hmap}, in contradiction. 
Thus, $H^+(\mathcal{S})$ has no endpoints and by
Proposition \ref{P:BKdif} must be (at least) of class $C^1$ at all points. $\Box$

{\em Outline of the proof of Theorem \ref{Th:Hcg}}
One can show that Lemma~\ref{l:hcom} can be adapted to the case 
of a compactly generated Cauchy horizon. The past-complete and
future-endless generators of $H^{+}(\mathcal{S}) $ are then contained 
in the compact set $C$ in Definition~\ref{d:cgen}~\cite{H92}.
One can also introduce the map $u_t$ described in the proof
of Lemma \ref{l:hexp} restricting it to the set $C$ where the integral formula
\ref{e:Hmap} holds as well. Suppose that the null convergence condition holds
then by past-completeness of the null generators of the horizon in the set $C$,
$\theta \geq 0$ and the right hand side of Eq. \ref{e:Hmap} cannot be negative. 
However if the compactly generated horizon is noncompact it will not lie completely 
in the compact set $C$ and consequently the left hand side of the Eq. \ref{e:Hmap}
has to be strictly negative. This gives a contradiction. $\Box$

{\em Proof of Theorem \ref{Th:Bch}}
Suppose that $H^{+}(\mathcal{S}) $ is a compact Cauchy horizon. 
By generic condition and the last part 
of Proposition \ref{l:hend} it follows that there will be an endless null generator
$\gamma$ of $H$ such that for some point $p$
on $\gamma$ \, $k^ck^dk_{[a}R_{b]cd[e}k_{f]} \neq 0$ where $k$ is a vector
tangent to $\gamma$ at $p$. By RNP equation and the null convergence condition
this is would mean that the expansion $\theta$ is non-zero somewhere
on $\lambda$ and this contradicts Lemma \ref{l:hexp}. $\Box$

By  Lemma~\ref{l:haend}, a compact Cauchy horizon cannot contain
even one generic generator. 
Thus we see that under Conditions \ref{c:conv} and \ref{c:gen} 
modulo certain differentiability assumptions compact Cauchy horizons 
are ruled out.

The above results are proved under the assumption of $C^2$ differentiablity
(modulo a set of measure zero) of the Cauchy horizon. 
Chru\'sciel et al. \cite{CDGH00} using the methods of geometric
measure theory have shown that it was possible to define expansion $\theta$
on non-differentiable event and Cauchy horizons. Consequently they were
able to give a proof of Hawking's black hole area theorem applicable
to any event horizon in an asymptotically flat smooth space-time.
They were also able to prove a remarkable theorem (Theorem 6.18 \cite{CDGH00})
that when $\theta$ vanishes on a Cauchy horizon and generators are endless 
then the horizon is as smooth
as the metric allows. If one were able to actually prove that $\theta$ 
is always zero on a compact Cauchy horizon and that generators are endless
(without almost smoothness
assumption) one would have a fundamental result that any compact Cauchy horizon
in a smooth space-time is smooth.

The results given above were stated for the case of compact and compactly
generated future Cauchy horizons. The time reverse versions hold for 
the case of compact and compactly generated past Cauchy horizons.

\section{Acknowledgments}

This work was supported by the Polish Committee for Scientific Research 
through grants 2~P03B~130~16 and 2~P03B~073~15.

\end{document}